\documentclass[11pt,twoside]{article}
\usepackage{asp2010}
\usepackage{natbib}

\resetcounters

\begin{document}

\title{On the Galactic Nova Progenitor Population}
\author{M.~J. Darnley$^1$, M.~F. Bode$^1$, D.~J. Harman$^1$, R.~A. Hounsell$^2$, U. Munari$^3$, V.~A.~R.~M. Ribeiro$^4$, F. Surina$^1$, R.~P. Williams$^1$ and S.~C. Williams$^1$
\affil{$^1$Astrophysics Research Institute, Liverpool John Moores University, Twelve Quays House, Egerton Wharf, Birkenhead, CH41~1LD, UK}
\affil{$^2$Space Telescope Science Institute, 3700 San Martin Drive, Baltimore, MD 21218, USA}
\affil{$^3$INAF Astronomical Observatory of Padova, via dell'Osservatorio, 36012, Asiago (VI), Italy}
\affil{$^4$Astrophysics, Cosmology and Gravity Centre, Department of Astronomy, University of Cape Town, Rondebosch 7701, South Africa}}

\begin{abstract}
Of the 350 or more known Galactic classical novae, only a handful of them, the recurrent novae, have been observed in outburst more than once.  At least eight of these recurrents are known to harbour evolved secondary stars, rather than the main sequence secondaries typical in classical novae.  Here we present a selection of the work and rationale that led to the proposal of a new nova classification scheme based not on the outburst properties but on the nature of the quiescent system.  Also outlined are the results of a photometric survey of a sample of quiescent Galactic novae, showing that the evolutionary state of the secondary can be easily determined and leading to a number of predictions.  We discuss the implications of these results, including their relevance to extragalactic work and the proposed link to type Ia supernovae.  We also present a summary of the work using the SMEI instrument to produce exquisite nova light-curves and confirmation of the pre-maximum halt.
\end{abstract}

\section{Introduction}

There are more than 350 known Galactic classical novae (CNe); of these only ten -- the recurrent novae (RNe) -- have been observed in outburst more than once.  Likewise, with around 900 CNe observed in the Andromeda Galaxy (M31) only a handful are RN candidates (see paper by A.~W.~Shafter, these proceedings).  The predicted recurrence times for CNe lie in the region of a few $\times10^{3}-10^{6}$~years, whereas the observed inter-outburst period for the RNe ranges from tens of years up to a hundred years.  This upper limit is clearly biased by the sparsity of historical records, particularly those predating the 20$^{\mathrm{th}}$ Century.

The secondary stars in CNe are typically main sequence stars (a notable exception is the intermediate polar GK~Per which hosts a sub-giant secondary), whereas the secondaries in the RNe are more evolved sub-giant or red giant stars (T~Pyx and IM~Nor being the exception to this rule).  The CN systems exhibit low mass transfer and accretion rates (leading to long inter-outburst times), contain white dwarfs with a range of masses and their outbursts evolve at all speed classes, showing either Fe~{\sc ii}, He/N or hybrid spectral classes.  RNe however, show high mass transfer and accretion rates and contain high mass white dwarfs (some close to the Chandrasekhar mass) both of which lead to the greatly reduced recurrence time.  RN outbursts (again with the exception of T~Pyx and IM~Nor) evolve quickly optically and exhibit He/N spectra.

Firstly, as an introduction, we present a brief summary of our work with the SMEI data archive in Section~\ref{SMEI}.  We then go on to discuss our work to-date on the Galactic nova progenitor population in a chronological manner.  In Section~\ref{M31N} we introduce our approach with a discussion of the extragalactic nova M31N~2007-12b.  In Section~\ref{RN} we discuss the extension of this approach to a number of Galactic systems, before leading on to the population study in Section~\ref{pop}.  Finally, in Sections~\ref{on} and \ref{exgal} we discuss some of the implications of our work.

\section{Solar Mass Ejection Imager (SMEI)}\label{SMEI}

The Solar Mass Ejection Imager (SMEI) was launched on board the {\it Coriolis} satellite on January 6$^{\mathrm{th}}$ 2003 and was in operation until the end of 2011.  The SMEI instrument consisted of three CCD cameras operating at $7000$~\AA\ with a FWHM $\sim3000$~\AA, each camera covered a $60^{\circ}\times3^{\circ}$ field of view.  SMEI monitored the entire sky with an uninterrupted cadence of 102 minutes, providing reliable photometry down to $\sim8^{\mathrm{th}}$ magnitude.  Work by \citet{2002AIPC..637..462S} indicated that as many as five of $34^{+15}_{-12}$ Galactic novae per year \citep{2006MNRAS.369..257D} would be luminous enough to be visible by SMEI.  The high cadence of SMEI coupled with its small Sun angle limitation makes it feasible to constrain the bright Galactic nova rate but also detect outbursts well before peak.

\citet{2010ApJ...724..480H} presented light curves of four particularly luminous novae within the SMEI archive; three CNe (KT~Eri, V598~Pup, V1280~Sco) and the 2006 outburst of the RN RS~Oph.  The light curves of these objects were unprecedented in their detail and completeness.  The SMEI data confirmed that the peak of both V598~Pup and KT~Eri had been missed by a considerable time despite both objects reaching naked eye visibility.  The 2011 outburst of the RN T~Pyx was also visible to SMEI (see paper by F.~Surina, these proceedings).

However, the most important SMEI result is the confirmation of the ``fabled'' pre-maximum halt \citep{2010ApJ...724..480H}.  The pre-max halt is clearly visible in the three fast novae from the catalogue, and may still be present in a different form in V1280~Sco.  The physical cause of this halt has now been investigated in detail by Y.~Hillman et~al.\ (these proceedings).

The SMEI database is still undergoing analysis to explore the population of bright (and nearby) Galactic novae, including possible ``missed'' systems.

\section{M31N 2007-12b}\label{M31N}

The nova M31N~2007-12b was discovered in outburst on 2007 December 9.53~UT by K.~Nishiyama and F.~Kabashima\footnote{\url{http://www.cfa.harvard.edu/iau/CBAT_M31.html}} within the Andromeda Galaxy (M31).  A broadband multi-colour photometric light-curve was obtained by the Liverpool Telescope \citep[LT;][]{2004SPIE.5489..679S} and optical spectroscopy by the Hobby Eberly Telescope both as part of a survey of Local Group extragalactic novae \citep{2011ApJ...734...12S,2012ApJ...752..156S}.  As reported by \citet{2009ApJ...705.1056B}, serendipitous {\it Swift} X-ray observations uncovered a super-soft source which appeared between 21 and 35 days after the optical maximum and had turned off by day 169.  Subsequent analysis indicated that the white dwarf in the system had a mass $\ge1.3M_{\odot}$ and that the optical light-curve and spectrum and the X-ray behaviour were consistent with that expected for a recurrent nova.  Archival {\it Hubble Space Telescope (HST)} ACS/WFC data were also available covering the region around the nova.  Astrometric and photometric analysis of the {\it HST} data revealed the progenitor system of the nova - a system containing a red giant secondary star \citep{2009ApJ...705.1056B}.

\citet{2011A&A...531A..22P} have since reported a determination of the white dwarf rotation period in the M31N~2007-12b system using a combination of {\it XMM-Newton} and {\it Chandra} observations.  These observations also indicated additional periodicity, possibly related to a (short) orbital period which would be at odds with the findings of \citet{2009ApJ...705.1056B}.  They conclude that M31N~2007-12b may be an intermediate polar (and would be the first detected extragalactically), 

\section{Recurrent Nova Candidates}\label{RN}

In recent years a growing number of CNe have been put forward as RN candidates, despite having only one observed outburst.  Such systems have typically been selected based upon the similarity of their X-ray and spectral behaviour to known recurrent novae (usually the sub-class prototypes RS~Oph and U~Sco).  The most significant of these systems is V2487~Oph for which, following a large archival search, a previous second outburst was identified \citep{2009AJ....138.1230P}.  Here we present a brief summary of our follow-up work for a number of these systems.

\subsection{V2491~Cygni}

Like V2487~Oph before it \citep{2002Sci...298..393H}, V2491~Cyg was also discovered in X-rays pre-outburst \citep{2009A&A...497L...5I}, being to-date the only two novae with such pre-outburst detections.  

Early post-outburst optical variation with a period of 0.1~days was reported by \citet{2008ATel.1514....1B} however this was unlikely to be related to the orbital period of the system which at this time was still obscured by the optically thick ejecta \citep{2011A&A...530A..70D}.  This variation is likely more akin to that reported by \citet{2010AJ....140..925S} seen early on in the 2010 outburst of U~Sco.

Subsequent optical photometric monitoring of the system at quiescence revealed optical flickering consistent with accretion but no periodicity $\le0.3$~days.  Optical and near-IR photometry of the progenitor indicated a system broadly consistent with a quiescent U~Sco; that is, a system containing a sub-giant secondary \citep{2011A&A...530A..70D}.

\subsection{V1721~Aquill\ae}

Nova V1721~Aql was discovered in September 2008 \citep{2008IAUC.8989....1Y} and reached $14^{\mathrm{th}}$ magnitude.  Early spectra indicated very high expansion velocities \citep{2008IAUC.8989....2H} and the system was initially mistaken for a supernova.  The initial spectra further indicated that the extinction towards this nova was very high.  \citet{2011A&A...530A..81H} reported on the full analysis of the photometric and spectral behaviour during outburst and upon the recovery of the progenitor system.

\citet{2011A&A...530A..81H} confirmed the high extinction $(A_{V}=11.6\pm0.2)$ and indicated that this was an extremely bright and fast nova very close to the Galactic plane at a distance of $2.2\pm0.6$~kpc.  Near-IR photometry of the progenitor system was consistent with both U~Sco and V2491~Cyg; again indicating the likelihood of a sub-giant secondary star.

\subsection{KT~Eridani}

Following the success of \citet{2009AJ....138.1230P} in recovering the progenitor system of V2487~Oph, and in uncovering an earlier missed outburst, a similar archival search was undertaken for KT~Eri.  \citet{2012A&A...537A..34J} conducted a search of the Harvard College Observatory archive which yielded the progenitor of KT~Eri and provided around 70 years worth of photometric coverage from 1890-1960.

Subsequent analysis of the archival data failed to reveal any previous outbursts although recurrence periods for the system of 41~years, 82~years or $>120$~years could not be ruled out.  The quiescent light-curve analysis also indicated a strong (orbital) periodicity of 737~days, the folded light-curve being reminiscent of reflection or eclipsing behaviour seen in symbiotic stars \citep[see their Figure~4]{2012A&A...537A..34J}.  For further discussion about the nature of this system see the paper by U.~Munari (these proceedings).

The archival progenitor photometry, coupled with 2MASS photometry at quiescence, the distance to the system ($\sim6.5$~kpc) and the above orbital period all point strongly at the secondary star being a red giant, albeit a low luminosity giant (unlike those found in RS~Oph and T~CrB).  This classification, \citep[one recently also made by][]{2012PASJ...64..120I}, indicates that there should be a continuum of red giant luminosities observed in nova systems, not just the high luminosity giants found so far.

\section{The Galactic Progenitor Population}\label{pop}

Traditionally, CNe have been classified by the properties of their outbursts, whereas RNe have tended to be classified by their quiescent state.  \citet{2012ApJ...746...61D} proposed a new, unifying, classification scheme for both CNe and RNe based solely on the evolutionary state of the secondary star.  This scheme introduced MS-novae, SG-novae and RG-novae; those systems containing main sequence, sub-giant and red giant secondaries respectively.

\begin{figure}
\plotone{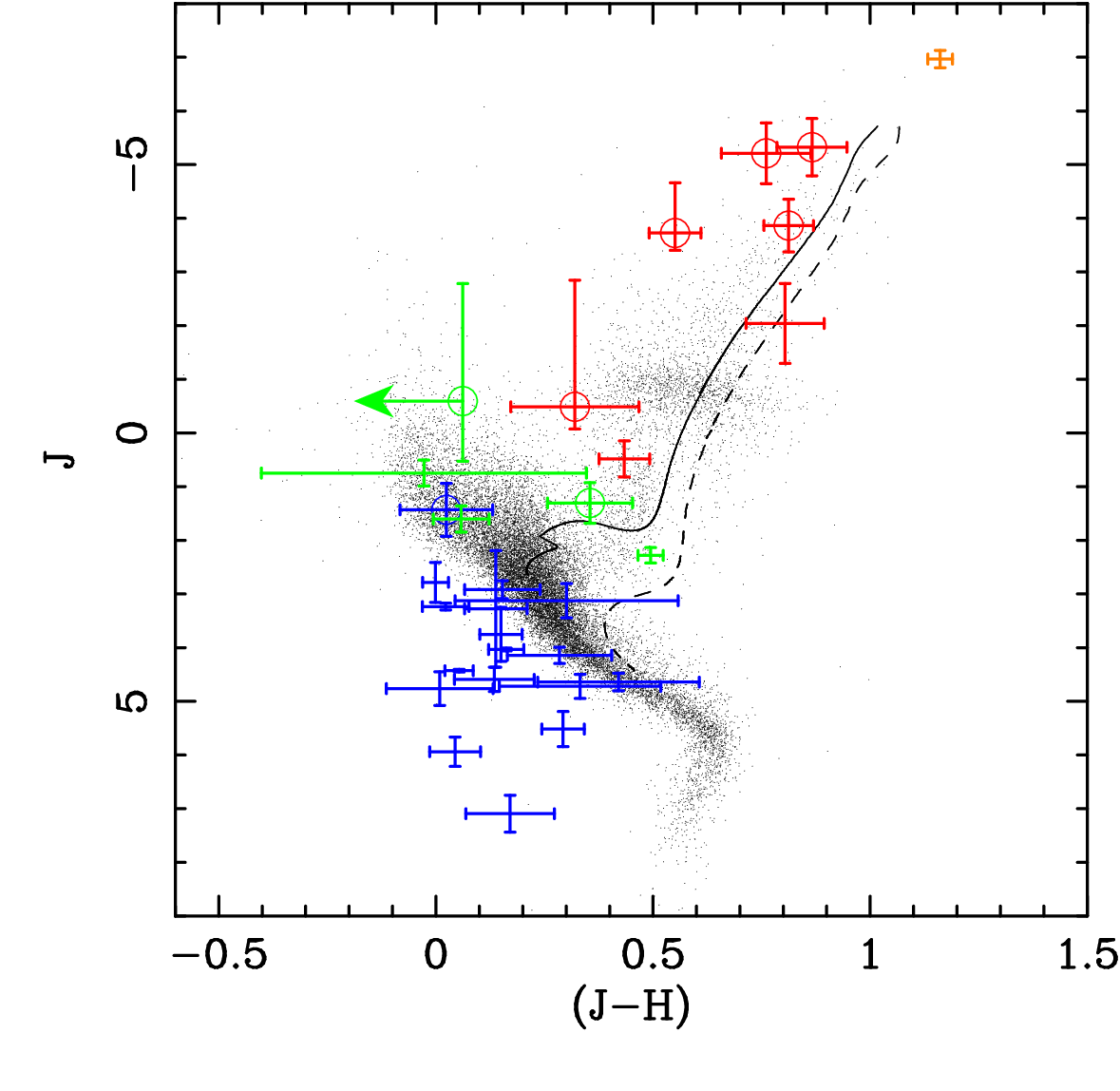}
\caption{Colour-magnitude diagram showing stars from the {\it Hipparcos} data set \citep{1997ESASP1200.....P} with parallax and photometric errors $<10\%$, near-IR photometry from the 2MASS catalogue \citep{2006AJ....131.1163S}.  The positions of the novae shown in \citet[see their Table~1]{2012ApJ...746...61D} are plotted.  Blue points represent MS-Novae, green and red points show members or candidates of the SG-Novae class and RG-Novae class, respectively, and the orange point shows the symbiotic/Mira nova V407 Cyg. The known recurrent novae in this sample have been identiﬁed by an additional circle. The black dashed line shows the evolutionary track of a 1~M$_{\odot}$ solar-like star, the solid line a 1.4~M$_{\odot}$ solar-like star \citep{2004ApJ...612..168P}.  Reproduced by permission of the AAS, from \citep{2012ApJ...746...61D}.\label{nearIR}}
\end{figure}

As can be seen in Figure~\ref{nearIR}, when the colour-magnitude positions of quiescent novae are plotted, the position of a particular system is a strong function of the evolutionary state of the secondary star.  When we focus on the near-IR, any dependence of the colour-magnitude position due to any accretion disk is minimised and in many cases negligible.  Systems harbouring RG secondaries are strongly correlated with the RGB.  Those containing MS secondaries are clustered blue-wards of the MS (due to accretion disk effects) but are less luminous than the local Galactic SG-branch.

\citet{2012ApJ...746...61D} highlighted a number of systems with evolved secondaries that have (to-date) only one observed outburst.  These include the SG-novae V1721~Aql and V2491~Cyg and the RG-novae KT~Eri, EU~Sct and M31N~2007-12b.  They also suggested the reclassification of the RN CI~Aql and DI~Lac to SG-novae and V2487~Oph to a RG-nova (similar to the KT~Eri system).  Spectroscopic follow-up observations are currently taking place to confirm these and other predictions.

\section{RG-Nova link to Type Ia Supernovae}\label{on}

A number of recent studies have cast some doubt on novae (specifically the ``recurrents'') being a significant SN~Ia progenitor channel.  These include \citet{2012Natur.481..164S} who failed to find a surviving secondary star around the remnant SNR~0509-67.5 in the LMC, and \citet{2011Natur.480..348L} who found that a RS~Oph/T~CrB type system was not the progenitor of SN~2011fe in M101.

\begin{figure}
\plotone{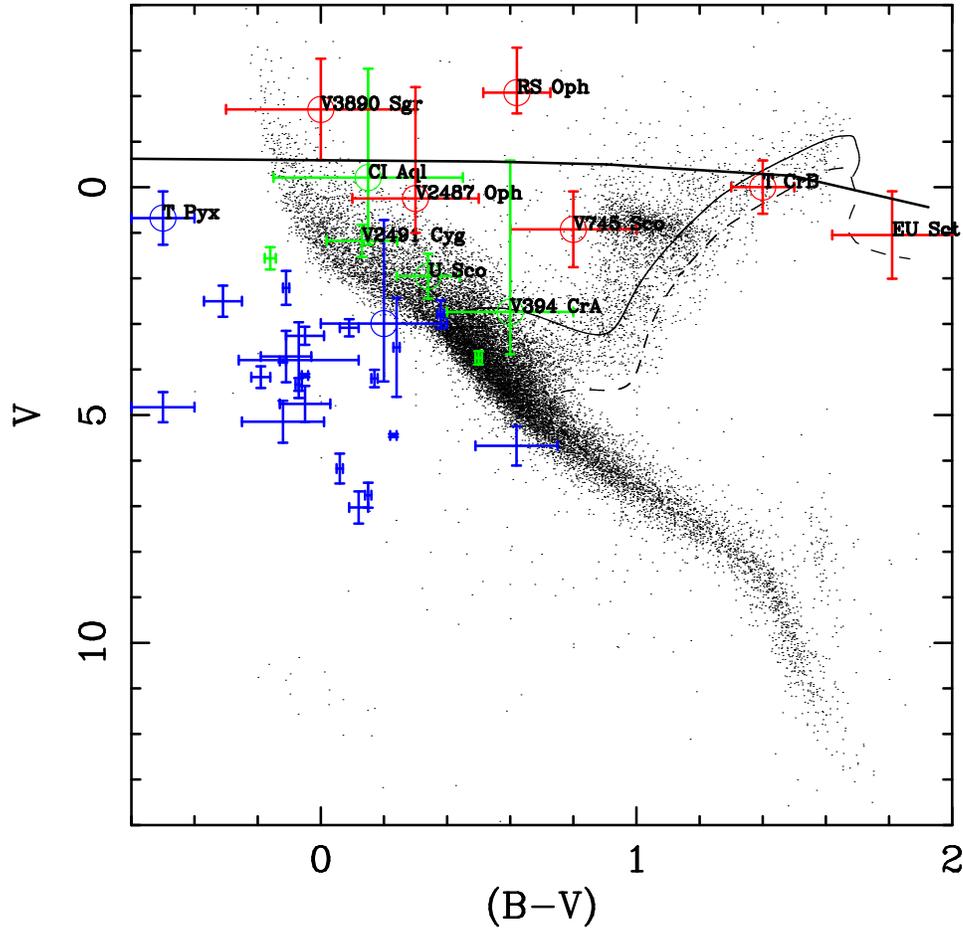}
\caption{As Figure~\ref{nearIR}, displaying optical photometry from the {\it Hipparcos} catalogue.  The thick solid black line indicates the $2\sigma$ detection level for progenitor systems of SN~2011fe in M101 from \citet{2011Natur.480..348L}.\label{li}}
\end{figure}

The \citet{2011Natur.480..348L} analysis was able to rule out progenitor systems containing ``luminous'' red giant stars (i.e.\ RG-novae) at the $2\sigma$ level.  However, this analysis did not take account of RG-novae with lower luminosity, less evolved secondary stars, e.g.\ the known recurrent nova V745~Sco, or KT~Eri.  Figure~\ref{li} presents a re-working of a plot from \citet[see their Figure~2]{2011Natur.480..348L}; here their $2\sigma$ level has been converted from temperature to $(B-V)$ colour using the information given in their Letter.  It is clear from Figure~2 that RG-novae with secondaries less luminous than that of RS~Oph could have been the progenitor system of SN~2011fe and have avoided detection by \citet{2011Natur.480..348L}.

\section{The Extragalactic Progenitor Population}\label{exgal}

The red giant population of Local Group galaxies, mainly M31, M32 and M33 is readily resolved by the {\it HST} and the larger ground based telescopes.  As such; as was successfully achieved for M31N~2007-12b, the progenitors of RG-novae are recoverable within the Local Group.  For example, {\it HST} WFC3/UVIS can in principle recover the entire RG-nova progenitor population within the Local Group, allowing us for the first time to directly measure the contribution these systems make to the entire nova population.  More exceptionally, {\it HST} WFC3/IR could recover some (luminous) SG-nova progenitor systems in regions where crowding is not too problematic (see the paper by S.~C.~Williams et~al., these proceedings, for further discussion in this area and preliminary results).

\bibliographystyle{asp2010}
\bibliography{darnley}

\end{document}